\documentclass[aps,pra,superscriptaddress,twocolumn]{revtex4}
\usepackage{bm}
\usepackage{graphicx,amsmath,latexsym,amssymb}
\usepackage{color}
\usepackage{dcolumn}

\begin{document}
\title{Two-atom interaction energies  with one atom in an excited state: van der Waals potentials vs. level shifts}
\author{M. Donaire}
\email{donaire@lkb.upmc.fr, mad37ster@gmail.com}
\affiliation{Laboratoire Kastler Brossel, UPMC-Sorbonnes Universit\'es, CNRS, ENS-PSL Research University,  Coll\`{e}ge de France, 4, place Jussieu, F-75252 Paris, France}

\begin{abstract}
I revisit the problem of the interaction between two dissimilar atoms with one atom in an excited state, recently addressed by the authors of Refs.\cite{Berman,Me,Milonni}, and for which precedent approaches have given conflicting results. In the first place, I discuss to what extent Refs.\cite{Berman}, \cite{Me} and \cite{Milonni} provide equivalent results. I show that the phase-shift rate of the two-atom wave function computed in Ref.\cite{Berman}, the van der Waals potential of the excited atom in Ref.\cite{Me} and the level shift of the excited atom in Ref.\cite{Milonni}  possess equivalent expressions in the quasistationary approximation. In addition, I show that the level shift of the ground state atom computed in Ref.\cite{Milonni} is equivalent to its van der Waals potential. A diagrammatic representation of all those quantities is provided. The equivalences among them are however not generic. In particular, it is found that for the case of the interaction between two identical atoms excited, the 
phase-shift rate and the van der Waals potentials differ. Concerning the conflicting results of previous approaches in regards to the spatial oscillation of the interactions, I conclude in agreement with Refs.\cite{Berman,Milonni} that they refer to different physical quantities.  The impacts of free-space dissipation and finite excitation rates on the dynamics of the potentials are analyzed. In contrast to Ref.\cite{Milonni}, the oscillatory versus monotonic spatial forms of the potentials of each atom are found not to be related to the reversible versus irreversible nature of the excitation transfer involved.
\end{abstract}
\maketitle

\section{Introduction}

Recently, several articles have addressed the problem of the van der Waals interaction between two dissimilar atoms, one of which is prepared in an excited state \cite{Berman,Me,Milonni,SafariPRL,Barcellona}. They all aim at explaining the spatial variation of the interaction at long interatomic separations, where  previous approaches have given conflicting results \cite{Sherkunov,Henkel,Passante,Power1995,Power1965,McLone}. Some of those articles put emphasis on the effect of  dissipation \cite{SafariPRL,Barcellona} and causality \cite{Me}, while others do on the physical meaning of the energies computed \cite{Berman,Milonni}. It has been found in Refs.\cite{Berman,Me,Milonni}, using time-dependent approaches, that free-space dissipation does not play an essential role in the computation. The calculation in Sec.\ref{TvdW} of the present paper so confirms.  Moreover, as firstly pointed out by Berman \cite{Berman} and then explicitly shown by Milonni and  Rafsanjani \cite{Milonni}, both the oscillatory and 
monotonic forms of the energy are possible as they refer to different physical quantities.
It is on this aspect that I concentrate here. The purpose of this paper is twofold. In the first place I discuss to what extent the quantities computed in Refs.\cite{Berman}, \cite{Me} and \cite{Milonni} are equivalent. Second, I analyze the effect of dissipation and finite excitation rates on the dynamics of the interaction.

The paper is organized as follows. After describing the fundamentals of the calculation and the nomenclature in Sec.\ref{Setup}, I explore in Sec.\ref{3papers} the physical meaning of the energies computed in Refs.\cite{Berman,Me,Milonni}  and discuss to what extent they are equivalent. In Sec.\ref{Rydberg} I present a case of interest where those quantities are manifestly different.  In Sec.\ref{TvdW} I analyze the impact of dissipation and finite excitation rates on the two-atom interaction potentials. I conclude with the Discussion in Sec.\ref{Disc}.

\section{Setup of the problem and fundamentals of the calculation}\label{Setup}

Let us consider two atoms, one of which is excited, located a distance $R$ apart. 
The excited atom is taken of type $A$ while the atom in its ground state is considered of a different type $B$. Both atoms are modeled by two-level systems of resonant frequencies $\omega_{A}$ and $\omega_{B}$ respectively, with respective linewidths $\Gamma_{A}$ and $\Gamma_{B}$ in free space.
Further, in order to preserve the perturbative nature of the calculation and to ensure the difference between the atomic species, the detuning
$\Delta_{AB}\equiv\omega_{A}-\omega_{B}$ is such that  $|\Delta_{AB}|\gg(\Gamma_{A}+\Gamma_{B})/2$ and $|\Delta_{AB}|\gg\langle W(T)\rangle/\hbar$,
with $W(T)$ being the interaction Hamiltonian at the time of observation, $T$. For the sake of simplicity, I will consider the quasiresonant approximation in most of the paper, setting $|\Delta_{AB}|\ll\omega_{A,B}$. This is the approximation considered  throughout Refs.\cite{Berman,Me} and in most of Ref.\cite{Milonni}. Eventually, I will consider also the quasistationary approximation in order to get rid off rapidly oscillating terms. This approximation has been applied throughout Ref.\cite{Milonni} by averaging over times  $T\gg2\pi/|\Delta_{AB}|$, throughout Ref.\cite{Berman} by considering the adiabatic excitation of atom $A$, and partially applied in Ref.\cite{Me} when considering the adiabatic switching of the interaction $W$. Further, radiative emission has been included in Refs.\cite{Berman,Milonni} in the Weisskopf-Wigner approximation. As shown in Ref.\cite{Me}, this inclusion is a higher order effect negligible  for observation times $T$ such that $\Gamma_{A,B}T\ll1$. Without much loss of 
generality, I will stick to the latter inequality throughout most of the paper until Sec.\ref{TvdW}. Finally, the state of the system at time 0 is $|\Psi(0)\rangle=|A_{+}\rangle\otimes|B_{-}\rangle\otimes|0_{\gamma}\rangle$, where  $(A,B)_{\pm}$ label the upper/lower internal states of the atoms $A$ and $B$ respectively and $|0_{\gamma}\rangle$ is the electromagnetic (EM) vacuum state. Implicitly, this  implies a sudden excitation of atom $A$. Again without much loss of generality, I will consider a sudden excitation until Sec.\ref{TvdW}.

At any given time $T>0$ the state of the two-atom-EM field system can be written as $|\Psi(T)\rangle=\mathbb{U}(T)|\Psi(0)\rangle$, where $\mathbb{U}(T)$ denotes the time propagator in the Schr\"odinger representation,
\begin{equation}
\mathbb{U}(T)=\mathcal{T}\exp{}\Bigl\{-i\hbar^{-1}\int_{0}^{T}\textrm{d}t\Bigl[ H_{A}+H_{B}+H_{EM}+W\Bigr]\Bigr\}.\nonumber
\end{equation}
In this equation $H_{A}+H_{B}$ is the free Hamiltonian of the internal atomic states, $\hbar\omega_{A}|A_{+}\rangle\langle A_{+}|+\hbar\omega_{B}|B_{+}\rangle\langle B_{+}|$, while the Hamiltonian of the free EM field is $H_{EM}=\sum_{\mathbf{k},\mathbf{\epsilon}}\hbar\omega(a^{\dagger}_{\mathbf{k},\mathbf{\epsilon}}a_{\mathbf{k},\mathbf{\epsilon}}+1/2)$,
where $\omega=ck$ is the photon frequency, and the operators $a^{\dagger}_{\mathbf{k},\mathbf{\epsilon}}$ and $a_{\mathbf{k},\mathbf{\epsilon}}$ are the creation and annihilation operators of photons with momentum $\hbar\mathbf{k}$ and polarization $\mathbf{\epsilon}$ respectively. Finally, the interaction Hamiltonian in the electric dipole approximation reads $W=W_{A}+W_{B}$, with $W_{A,B}=-\mathbf{d}_{A,B}\cdot\mathbf{E}(\mathbf{R}_{A,B})$. In this expression $\mathbf{d}_{A,B}$ are the electric dipole operators of each atom and  $\mathbf{E}(\mathbf{R}_{A,B})$ is the electric field operator evaluated at the position of each atom, which   can be written in the usual manner as a sum over normal modes,
\begin{eqnarray}\label{AQ}
\mathbf{E}(\mathbf{R}_{A,B})&=&\sum_{\mathbf{k}}\mathbf{E}^{(-)}_{\mathbf{k}}(\mathbf{R}_{A,B})+\mathbf{E}^{(+)}_{\mathbf{k}}(\mathbf{R}_{A,B})\nonumber\\
&=&i\sum_{\mathbf{k},\mathbf{\epsilon}}\sqrt{\frac{\hbar ck}{2\mathcal{V}\epsilon_{0}}}
[\mathbf{\epsilon}a_{\mathbf{k},\epsilon}e^{i\mathbf{k}\cdot\mathbf{R}_{A,B}}-\mathbf{\epsilon}^{*}a^{\dagger}_{\mathbf{k},\epsilon}e^{-i\mathbf{k}\cdot\mathbf{R}_{A,B}}],\nonumber
\end{eqnarray}
where $\mathcal{V}$ is a generic volume and $\mathbf{E}^{(\mp)}_{\mathbf{k}}$ denote the annihilation/creation electric field operators of photons of momentum $\hbar\mathbf{k}$, respectively. 

Next, considering $W$ as a perturbation to the free Hamiltonians, the unperturbed time propagator for atom and free photon states is $\mathbb{U}_{0}(t)=\exp{[-i\hbar^{-1}(H_{A}+H_{B}+H_{EM})t]}$. In terms of $W$ and $\mathbb{U}_{0}$, $\mathbb{U}(T)$ admits an expansion in powers of  $W$ which can be developed out of the time-ordered exponential equation,
\begin{equation}\label{U}
\mathbb{U}(T)=\mathbb{U}_{0}(T)\:\mathcal{T}\exp\int_{0}^{T}(-i/\hbar)\mathbb{U}_{0}^{\dagger}(t)\:W\:\mathbb{U}_{0}(t)\textrm{d}t.
\end{equation}
Finally and for further purposes, I denote the term of order $W^{n}$ in the corresponding series by $\delta\mathbb{U}^{(n)}$ and write $\mathbb{U}(T)=\mathbb{U}_{0}(T)+\sum_{n=1}^{\infty}\delta\mathbb{U}^{(n)}(T)$. As an example, $\delta\mathbb{U}^{(2)}$ reads
\begin{equation}
\delta\mathbb{U}^{(2)}(T)=-\hbar^{2}\int_{0}^{T}\textrm{d}t\int_{0}^{t}\textrm{d}t'\mathbb{U}_{0}(T-t)W\mathbb{U}_{0}(t-t')W\mathbb{U}_{0}(t').
\end{equation}

\section{Comparison between the energies computed in Refs.\cite{Berman}, \cite{Me} and \cite{Milonni}}\label{3papers}

The time-dependent approaches of Refs.\cite{Berman}, \cite{Me} and \cite{Milonni} are equivalent as they all compute interaction energies up to fourth order in $W$, make use of equivalent approximations and consider the same initial state $|\Psi(0)\rangle$. Therefore,
the quantities there computed can be compared to each other when expressed in the same representation. To this end, I first express the quantities of Refs.\cite{Berman,Me,Milonni} in Schr\"odinger's representation, using the nomenclature of Sec.\ref{Setup}. Next, I investigate their physical meanings and discuss the relationship between them.

\subsection{The van der Waals potentials in Ref.\cite{Me}}\label{Mer}
Let us start by defining the time-dependent quadratic vdW potential (vdW potential, in brief) as the effective
potential energy from which the vdW force on each atom is derived upon application of the classical operators $-\nabla_{\overline{\mathbf{R}}_{A}}$ and $-\nabla_{\overline{\mathbf{R}}_{B}}$, respectively, with  $\overline{\mathbf{R}}_{A,B}=\langle\mathbf{R}_{A,B}\rangle$. That is, considering both atoms at rest, the vdW force on each atom  at  order $W_{A}^{2}W_{B}^{2}$ is
\begin{align}
\langle\mathbf{F}_{A,B}(T)\rangle&=\partial_{T}\langle\mathbf{Q}_{A,B}(T)\rangle\nonumber\\
&=-i\hbar\partial_{T}\langle\Psi(0)|\mathbb{U}^{\dagger}(T)\mathbf{\nabla}_{\mathbf{R}_{A,B}}\mathbb{U}(T)|\Psi(0)\rangle\nonumber\\
&=-\langle \mathbf{\nabla}_{\mathbf{R}_{A,B}}W_{A,B}(T,\mathbf{R}_{A,B})\rangle\nonumber\\&=\frac{-1}{2}\mathbf{\nabla}_{\overline{\mathbf{R}}_{A,B}}\langle W_{A,B}(T,\overline{\mathbf{R}}_{A,B})\rangle,\label{Force}
\end{align}
where $\mathbf{Q}_{A,B}$ are the kinetic momenta of the centers of mass of each atom. In the last line, by replacing the gradients with respect to the quantum variables $\mathbf{R}_{A,B}$ with the gradients with respect to the classical variables $\overline{\mathbf{R}}_{A,B}$, I am assuming that quantum fluctuations are negligible over  $\overline{\mathbf{R}}_{A,B}$. The factor 1/2 in front of $\mathbf{\nabla}_{\overline{\mathbf{R}}_{A,B}}$ comes from the fact that  $|\Psi(T)\rangle\langle\Psi(T)|$ is already of orders $W_{A}W^{2}_{B}$ and $W_{B}W^{2}_{A}$ in the calculations of $\langle\mathbf{F}_{A}(T)\rangle$ and $\langle\mathbf{F}_{B}(T)\rangle$, respectively. Physically, the factor 1/2 is a consequence of the fact that, for an induced atomic dipole, half of $\langle W_{A,B}(T)\rangle$ contributes to the polarization of the atomic states of $A,B$ respectively. Therefore, the vdW potentials are $\langle W_{A,B}(T)\rangle/2$. $\langle W_{A}(T)\rangle/2$ is indeed the quantity computed in Ref.\cite{Me} in 
the quasiresonant approximation, where a factor $1/2$ was missing on the left
hand side of Eq.(2) and thereafter.

\begin{figure}[h]
\includegraphics[height=10.7cm,width=9.0cm,clip]{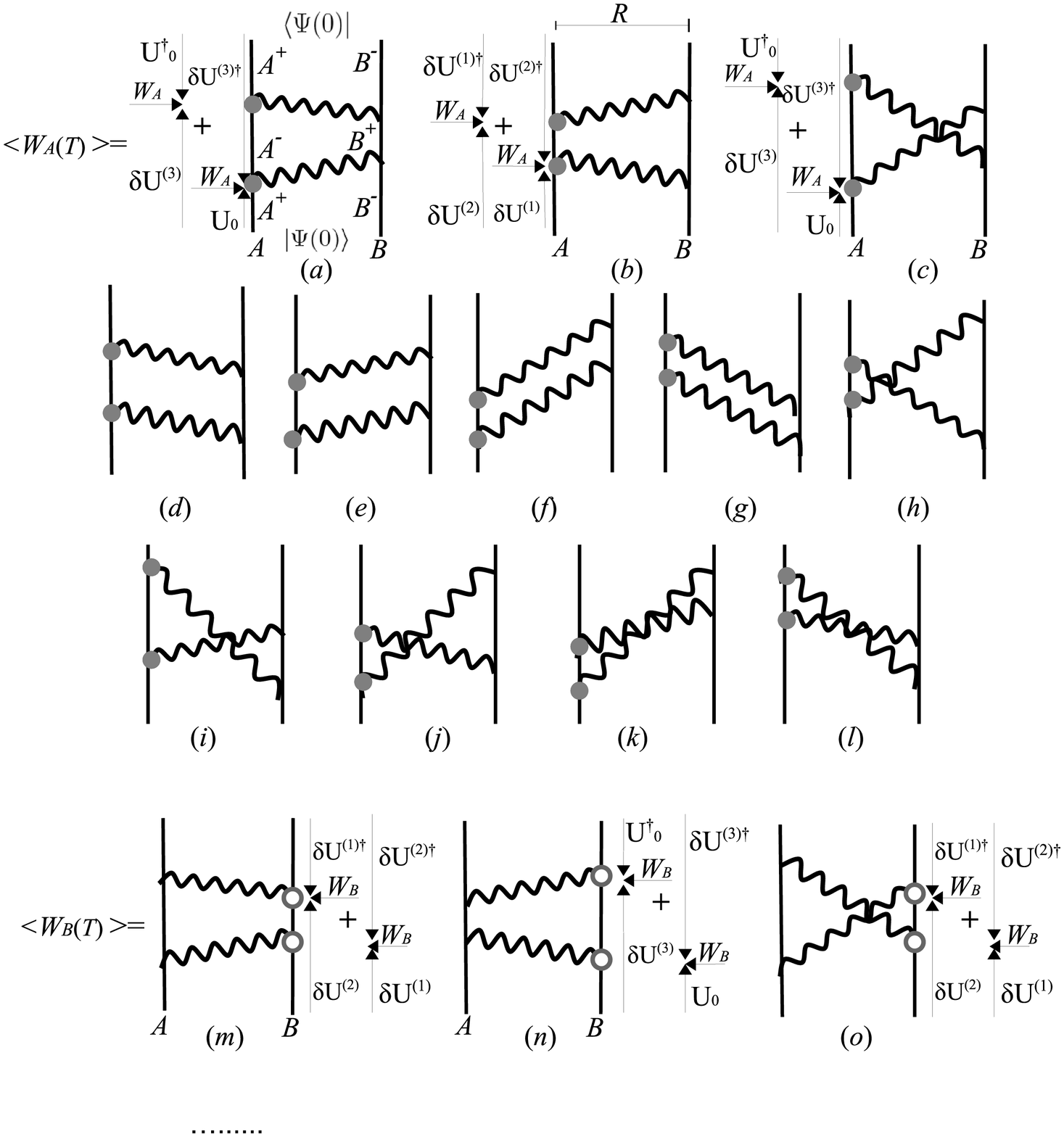}
\caption{Diagrammatic representation of (twice) the vdW potentials, $\langle W_{A}(T)\rangle$ and $\langle W_{B}(T)\rangle$. Twelve diagrams contribute to each one, but only three of them are depicted for $\langle W_{B}(T)\rangle$ for brevity.
Thick straight lines stand for propagators of atomic states, while wavy lines do for photon propagators. The atoms are separated by a distance $R$ along
the horizontal direction, whereas time runs along the vertical. The grey circles on the left of each diagram stand for the insertion of a Schr\"odinger operator $W_{A}$, while white circles
on the right denote the application of an operator $W_{B}$. Each diagram contributes with two terms to each potential, as explicitly written in
diagrams $(a)-(c)$ and $(m)-(o)$. In those diagrams, either an operator $W_{A}$ or $W_{B}$ is sandwiched between two time propagators (depicted by arrows) which evolve
the intial states $|\psi(0)\rangle$ and $\langle\psi(0)|$ towards the observation time $T$, at which either $W_{A}$ or $W_{B}$ apply.}\label{figure1}
\end{figure}

Up to order $W_{A}^{2}W_{B}^{2}$,  twelve are the diagrams which contribute to $\langle W_{A}(T)\rangle$ and depend on $\mathbf{R}_{A}$, hence observable through the measurement of the vdW force $\langle\mathbf{F}_{A}(T)\rangle$. They are depicted in Figs.\ref{figure1}$(a)-(l)$. Analogous diagrams hold for $\langle W_{B}(T)\rangle$, three of which have been represented in Figs.\ref{figure1}$(m)-(o)$.  In the quasiresonant approximation ($qr$), $|\Delta_{AB}|/\omega_{A,B}\ll1$, the dominant contributions to each potential come from diagrams $(a)$ and $(m)$ of Fig.\ref{figure1}, respectively. The calculation of $\langle W_{A}(T)\rangle/2$ was already explained in Ref.\cite{Me}. As for $\langle W_{B}(T)\rangle/2$, its integral equation is given in Eq.(\ref{ECUACIONFORWBT}) of the Appendix for a sudden excitation. In the far field, $k_{A,B}R\gg1$, their expressions are, respectively,
\begin{align}
\langle W_{A}(T)\rangle^{qr}/2&\simeq\Theta(T-2R/c)\frac{\mathcal{U}_{ijpq}}{R^{2}}\alpha^{ij}\alpha^{pq}[k_{A}^{4}\cos{(2k_{A}R)}\nonumber\\
&-k_{B}^{4}\cos{(2k_{B}R+\Delta_{AB}T)}],\label{WA}
\end{align}
\begin{align}
\langle W_{B}(T)\rangle^{qr}/2&\simeq\Theta(T-R/c)\frac{\mathcal{U}_{ijpq}}{R^{2}}\alpha^{ij}\alpha^{pq}\bigl[k_{A}^{4}\nonumber\\
&-k_{B}^{2}k_{A}^{2}\cos{[\Delta_{AB}(T-R/c)]}\bigr],\label{WB}
\end{align}
where $\mathcal{U}_{ijpq}=\mu^{A}_{i}\mu^{A}_{q}\mu^{B}_{j}\mu^{B}_{p}/[(4\pi\epsilon_{0})^{2}\hbar\Delta_{AB}]$,
$\mu^{A}_{i}=\langle A_{-}|d_{A,i}|A_{+}\rangle$, $\mu^{B}_{j}=\langle B_{-}|d_{B,j}|B_{+}\rangle$ and $\alpha^{ij}=\delta^{ij}-R^{i}R^{j}/R^{2}$, with $\mathbf{R}=\overline{\mathbf{R}}_{A}-\overline{\mathbf{R}}_{B}$. We observe that the time-dependent terms of both potentials oscillate with frequency $\Delta_{AB}$. However, while the time-independent term of  $\langle W_{A}(T)\rangle^{qr}/2$  oscillates in space with frequency $2k_{A}$, that of $\langle W_{B}(T)\rangle^{qr}/2$ has a monotonic form.

Lastly,  the dynamical vdW potentials can be averaged in time for $T\gg|\Delta_{AB}^{-1}|$  in order to obtain quasistationary values. By doing so, beyond the far field limit,  one obtains
\begin{align}
\langle W_{A}/2\rangle^{qr}_{T}&=(4\pi k^{2}_{A})^{2}\mathcal{U}^{ijpq}\nonumber\\&\times\Bigl[\textrm{Re}\:G_{ij}^{(0)}(\mathbf{R},\omega_{A})\textrm{Re}\:G_{pq}^{(0)}(\mathbf{R},\omega_{A})\nonumber\\
&-\textrm{Im}\:G_{ij}^{(0)}(\mathbf{R},\omega_{A})\textrm{Im}\:G_{pq}^{(0)}(\mathbf{R},\omega_{A})\Bigr]\nonumber\\
&=\frac{\mathcal{U}_{ijpq}}{R^{6}}[\beta^{ij}\beta^{pq}-k_{A}^{2}R^{2}(\beta^{ij}\beta^{pq}+2\alpha^{ij}\beta^{pq})\nonumber\\
&+k_{A}^{4}R^{4}\alpha^{ij}\alpha^{pq}]\cos{(2k_{A}R)}+\frac{2\mathcal{U}_{ijpq}}{R^{5}}k_{A}[\beta^{ij}\beta^{pq}\nonumber\\
&-k_{A}^{2}R^{2}\alpha^{ij}\beta^{pq}]\sin{(2k_{A}R)},\label{WAT}
\end{align}
\begin{align}
\langle W_{B}/2\rangle^{qr}_{T}&=(4\pi k^{2}_{A})^{2}\mathcal{U}^{ijpq}\nonumber\\&\times\Bigl[\textrm{Re}\:G_{ij}^{(0)}(\mathbf{R},\omega_{A})\textrm{Re}\:G_{pq}^{(0)}(\mathbf{R},\omega_{A})\nonumber\\
&+\textrm{Im}\:G_{ij}^{(0)}(\mathbf{R},\omega_{A})\textrm{Im}\:G_{pq}^{(0)}(\mathbf{R},\omega_{A})\Bigr]\nonumber\\
&=\mathcal{U}_{ijpq}[\beta^{ij}\beta^{pq}/R^{6}+(\beta^{ij}\beta^{pq}-2\alpha^{ij}\beta^{pq})k_{A}^{2}/R^{4}\nonumber\\&+\alpha^{ij}\alpha^{pq}k^{4}/R^{2}],\label{WBT}
\end{align}
where $\langle\mathcal{O}\rangle_{T}$ denotes the quasistationary expectation value of a quantum operator $\mathcal{O}$, and $\mathbb{G}^{(0)}(\mathbf{R},\omega)$ is the dyadic Green's function of the electric field induced at $\mathbf{R}$ by an electric dipole of frequency $\omega=ck$ in free space. It reads
\footnote{In Ref.\cite{Milonni}, the dyadic $F$ defined in Eq.(22) equals $4\pi k_{A}^{2}\hbar_{-1}|\mu_{A}||\mu_{B}|e^{-ik_{A}R}\mathbb{G}^{(0)}(\mathbf{R},\omega_{A})$. In the Supplemental Material of Ref.\cite{Me}, the dyadic $\mathcal{F}(kR)$ defined in Eq.(SM1) equals $4\pi k^{-1}\textrm{Im}\mathbb{G}^{(0)}(\mathbf{R},\omega)$.}
\begin{equation}
\mathbb{G}^{(0)}(\mathbf{R},\omega)=\frac{k\:e^{ikR}}{4\pi}[\alpha/kR+i\beta/(kR)^{2}-\beta/(kR)^{3}],
\end{equation}
where the tensors $\alpha$ and $\beta$ read $\alpha=\mathbb{I}-\mathbf{R}\mathbf{R}/R^{2}$,  $\beta=\mathbb{I}-3\mathbf{R}\mathbf{R}/R^{2}$.

In Eqs.(\ref{WAT}) and (\ref{WBT}) I have firstly expressed  the vdW potentials in terms of  $\mathbb{G}^{(0)}$ in order to show that the difference between them finds in the opposite contribution of  Im$^{2}\mathbb{G}^{(0)}$.  As a result,  $\langle W_{A}/2\rangle^{qr}_{T}$ presents spatial oscillations while $\langle W_{B}/2\rangle^{qr}_{T}$ is monotonic. These are the results obtained by Milonni \& Rafsanjani \cite{Milonni} for analogous quantities --see below.\\

Berman \cite{Berman} and Milonni \& Rafsanjani \cite{Milonni} have identified  energy level shifts  from the expressions of the expectation values of certain atomic operators.  In particular, Berman has computed, in the interaction representation, the time derivative of the probability amplitude of finding atom $A$ in the state $A_{+}$ at time $T$, $\dot{b}_{A}(T)$. On the other hand, Milonni \& Rafsanjani \cite{Milonni} have computed, in Heisenberg's representation, the time derivative of the expectation value of the two-state lowering operators, $\langle\dot{\sigma}_{A,B}(T)\rangle$, with $\sigma_{A}=|A_{-}\rangle\langle A_{+}|$, $\sigma_{B}=|B_{-}\rangle\langle B_{+}|$. In both cases, in order to identify the corresponding level shifts, the authors have taken a quasistationary approximation averaging in time their equations for $T\gg|\Delta_{AB}^{-1}|$  (or, equivalently, by assuming the adiabatic excitation of atom $A$ \cite{Berman}). I study these quantities in the following.

\subsection{The phase shift rate of the two-atom wave function in Ref.\cite{Berman}}
In this section I analyze the expression for the energy shift $\delta\mathcal{E}$ computed in Ref.\cite{Berman} and I compare it to the vdW potentials of the previous section. In terms  of Schr\"odinger's propagators, the probability amplitude of finding atom $A$ excited at time $T$ reads, in the interaction picture,
\begin{align}
b_{A}(T)&=\partial_{T}\textrm{Tr}_{B}\langle A_{+}|\otimes\langle B|\otimes\langle0_{\gamma}|\mathbb{U}_{0}^{\dagger}(T)\mathbb{U}(T)|\Psi(0)\rangle\nonumber\\
&=\partial_{T}\langle\Psi(0)|\mathbb{U}_{0}^{\dagger}(T)\mathbb{U}(T)|\Psi(0)\rangle.\label{bA}
\end{align}
In this equation $\mathbb{U}_{0}^{\dagger}(T)\mathbb{U}(T)$ is the time propagator in the interaction representation, the trace in the first line is taken over the atomic states of  $B$, $|B_{\pm}\rangle$, and $|B_{+}\rangle$ has been dropped in the second line because its contribution vanishes. From Eq.(\ref{bA}) we read that $b_{A}(T)$ is the wave function of the two-atom state in the interaction representation ($I$), $\langle\Psi(0)|\Psi_{I}(T)\rangle$. Berman identifies $\delta\mathcal{E}$ with the real part of $i\hbar\langle\dot{b}_{A}\rangle_{T}$.
This identification lies in the assumption that, in the quasistationary approximation, $b_{A}(T)$ can be approximated by $e^{-i\hbar^{-1}T(\delta\mathcal{E}-i\Gamma_{A}/2)}$, which implies implicitly that $|\delta\mathcal{E}/\hbar\Delta_{AB}|\ll1$ in order to neglect the population of the state $|A_{-}\rangle\otimes|B_{+}\rangle$.
This being the case,  $T\delta\mathcal{E}/\hbar$ is the phase-shift accumulated by the two-atom wave function since the time atom $A$ was excited. Correspondingly, $\delta\mathcal{E}/\hbar$ is the rate of phase-shift. Lastly, given that  $\delta\mathcal{E}$ is already of order $W^{4}$, at the lowest order in $W$ it holds that\footnote{In strict sense, it holds $\delta\mathcal{E}\simeq\textrm{Re}\{i\hbar\langle b^{*}_{A}\rangle_{T}\langle\dot{b}_{A}\rangle_{T}\}$ instead. However, since $\langle\dot{b}_{A}\rangle_{T}$ is already of order $W_{A}^{2}W_{B}^{2}$, we take $\langle b^{*}_{A}\rangle_{T}\simeq1$ at the lowest order.} $\delta\mathcal{E}\simeq\textrm{Re}\{i\hbar\langle\dot{b}_{A}\rangle_{T}\}$.

\begin{figure}[h]
\includegraphics[height=9.6cm,width=8.9cm,clip]{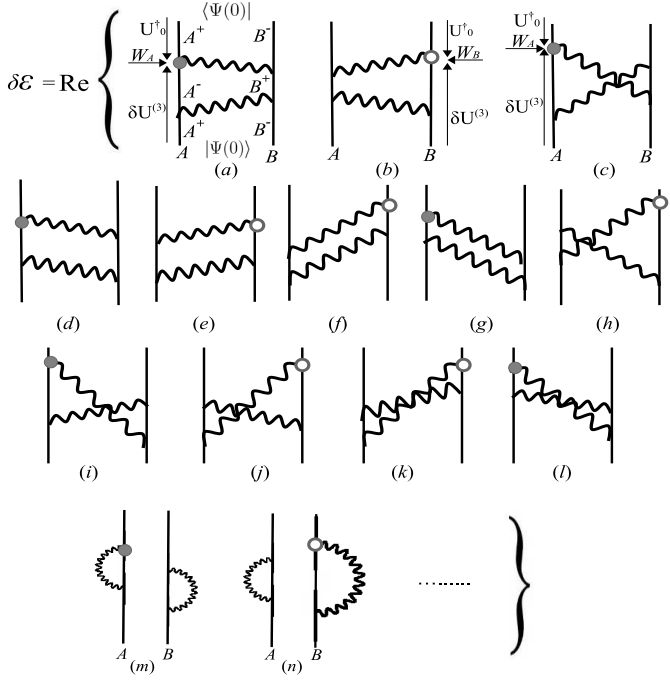}
\caption{Diagrammatic representation of the phase-shift rate of the two-atom wave function, $\delta\mathcal{E}$. Beside the first twelve $R$-dependent diagrams which contribute to the interaction potentials, there are $R$-independent diagrams whose contributions are negligible in good approximation. Only two of the latter
are depicted for brevity. Each diagram contributes with an only term to $\delta\mathcal{E}$, as explicitly indicated in
diagrams $(a)-(c)$. Reading those diagrams from the bottom up, it is the last operator, either $W_{A}$ or $W_{B}$, that is sandwiched between
the free propagator $\mathbb{U}^{\dagger}_{0}(T)$ on the left and a term of $\delta\mathbb{U}^{(3)}(T)$ on the right. The former propagates the state $\langle\psi(0)|$, while the
latter propagates $|\psi(0)\rangle$ towards the observation time $T$ at which either $W_{A}$ or $W_{B}$ apply.}\label{figure2}
\end{figure}

In order to find the relation of $\delta\mathcal{E}$ with the vdW potentials, I first retain the time dependence of $\textrm{Re}\{i\hbar\dot{b}_{A}(T)\}$ before averaging in time. Differentiating Eq.(\ref{bA}) with respect to $T$ and using Eq.(\ref{U}) one gets
\begin{equation}
\textrm{Re}\{i\hbar\dot{b}_{A}(T)\}=\textrm{Re}\{\langle\Psi(0)|\mathbb{U}_{0}^{\dagger}(T)(W_{A}+W_{B})\mathbb{U}(T)|\Psi(0)\rangle\}.\label{dEA}
\end{equation}
The relation of $\textrm{Re}\{i\hbar\dot{b}_{A}(T)\}$ to the vdW potentials is better shown diagrammatically. In Fig.\ref{figure2} I have depicted the diagrams which contribute to $\textrm{Re}\{i\hbar\dot{b}_{A}(T)\}$ up to order $W^{4}$. As can be readily seen, only the  contributions of the $R$-dependent diagrams $(a),(c),(d),(g),(i),(l)$ coincide with half of the contributions of the same diagrams in Fig.\ref{figure1} for $\langle W_{A}(T)\rangle/2$, while the contributions of the remaining six $R$-dependent diagrams pertain to $\langle W_{B}(T)\rangle/2$. Therefore, generally, it holds that $\textrm{Re}\{i\hbar\dot{b}_{A}(T)\}\neq\langle W_{A}(T)\rangle/2$, $\langle W_{B}(T)\rangle/2$.

Let us first restrict to the quasiresonant approximation, in which the dominant contribution comes from diagram \ref{figure2}$(a)$. Due to its up-down symmetry and to the fact that, when read from bottom up, the last  photon is annihilated at atom $A$, its contributions to  $\textrm{Re}\{i\hbar\dot{b}_{A}(T)\}$ and $\langle W_{A}(T)\rangle/2$ coincide. Hence, the real part of the r.h.s. of Eq.(33) in Ref.\cite{Berman} for $\textrm{Re}\{i\hbar\dot{b}_{A}(T)\}$ equals the r.h.s. of Eqs.(2,SM2) in Ref.\cite{Me} for  $\langle W_{A}(T)\rangle/2$ (with $\Gamma_{A,B}T\rightarrow0$).

Next, in the quasistationary approximation and beyond the quasiresonant approximation,  the discrepancy between $\delta\mathcal{E}$ and $\langle W_{A}/2\rangle_{T}$  may be only caused by the resonant diagrams $(a)$, $(c)$ of Figs.\ref{figure1} and \ref{figure2}. However, again because of the up-down symmetry of the diagrams $(a)$, $(c)$, they both give identical contributions to $\delta\mathcal{E}$ and $\langle W_{A}/2\rangle_{T}$. In particular, they yield the frequency poles --see Eq.(SM5) in Ref.\cite{Me} and Eq.(\ref{diagc}) in the Appendix--
\begin{align}
&c/[\/\Delta_{AB}(k-k_{A}-i\eta/c)(k'-k_{A}-i\eta/c)]\quad\textrm{and}\label{poleA}\\
&-c/[(k+k'-\/\Delta_{AB})(k-k_{A}-i\eta/c)(k'-k_{A}-i\eta/c)],\nonumber
\end{align}
respectively, with $\eta\rightarrow0^{+}$. The contribution of the poles provided by diagram $(c)$ is the same as that of diagram $(a)$ in Eq.(\ref{WAT}) but for the replacement of $1/\Delta_{AB}$ with $-1/(\omega_{A}+\omega_{B})$ \footnote{This was also noted by the authors of Ref.\cite{Milonni} in the calculation of $\delta\omega_{A}$ --see next Section. There is however a minus sign missing in the second term on the r.h.s. of Eq.(36) in Ref.\cite{Milonni}. Hence, the l.h.s. should read instead $2\omega_{B}/(\omega_{A}^{2}-\omega_{B}^{2})$.} \footnote{In contrast to the frequency integrals of diagram \ref{figure2}$(a)$ [also of diagram \ref{figure3}$(a)$ later], the lower bound of the frequency integrals which contain the poles of diagram \ref{figure2}$(c)$ [also of diagram \ref{figure3}$(d)$ later] are not to be extended to $-\infty$.}.

From this analysis I conclude that the results of Refs.\cite{Berman} and \cite{Me} are fully equivalent in the quasistationary approximation and beyond the quasiresonant condition, even though they refer to different physical quantities.

\subsection{The single-atom level shifts in Ref.\cite{Milonni}}
In this section I analyze the expressions for the level shifts computed in Ref.\cite{Milonni} in order to explain their relationship with the vdW potentials and the phase-shift rate described in the previous Sections. As already mentioned, Milonni and Rafsanjani \cite{Milonni} have identified the single-atom frequency shifts  $\delta\omega_{A,B}$ in the equations of motion of the expectation values of the two-state lowering operators, $\langle\sigma_{A,B}(T)\rangle$. Neglecting radiation reaction terms in the limit $\Gamma_{A,B}T\ll1$,  the  Heisenberg equations for   $\langle\sigma_{A,B}(T)\rangle$ read [cf. Eq.(2) in Ref.\cite{Milonni}], in terms of Schr\"odinger's (first two lines on the r.h.s.) and Heisenberg's operators (last two lines on the r.h.s.),
\begin{align}\label{dotsigmaB}
\langle\dot{\sigma}_{A,B}(T)\rangle&=-i\omega_{A,B}\langle\Psi(0)|\mathbb{U}^{\dagger}(T)\sigma_{A,B}\mathbb{U}(T)|\Psi(0)\rangle\nonumber\\
&+i\hbar^{-1}\langle\Psi(0)|\mathbb{U}^{\dagger}(T)[W_{A,B}\sigma_{A,B}\mathbb{U}(T)|\Psi(0)\rangle\nonumber\\
&=-i\omega_{A,B}\langle\sigma_{A,B}(T)\rangle+i\hbar^{-1}[\langle W_{A,B}(T)\sigma_{A,B}(T)\rangle\nonumber\\
&-\langle\sigma_{A,B}(T)W_{A,B}(T)\rangle].
\end{align}
The identifications of $\delta\omega_{A,B}$ rest on the assumption that, under quasistationary conditions, Eq.(\ref{dotsigmaB}) can be written as
\begin{equation}\label{dotsigmaBstat}
\langle\dot{\sigma}_{A,B}\rangle_{T}\simeq-i(\omega_{A,B}+\delta\omega_{A,B})\langle\sigma_{A,B}\rangle_{T}+\langle....\rangle,
\end{equation}
where $\langle....\rangle$ denotes terms which do not depend on $\langle\sigma_{A,B}\rangle_{T}$ in either case and dissipative terms have been again discarded. The quantities of interest are the second terms on the r.h.s. of Eq.(\ref{dotsigmaB}), which I will refer to as $\delta\langle\dot{\sigma}_{A,B}(T)\rangle$.
In the following, I will restrict its calculation to the approximations considered in Ref.\cite{Milonni}, which are indeed equivalent to the approximations of Refs.\cite{Berman,Me}. That is, the authors there  consider  the contribution of resonant photons to $\delta\langle\dot{\sigma}_{A,B}(T)\rangle$  in the rotating wave approximation --which is equivalent in this context to the quasiresonant approximation.

\begin{figure}[h]
\includegraphics[height=6.5cm,width=8.5cm,clip]{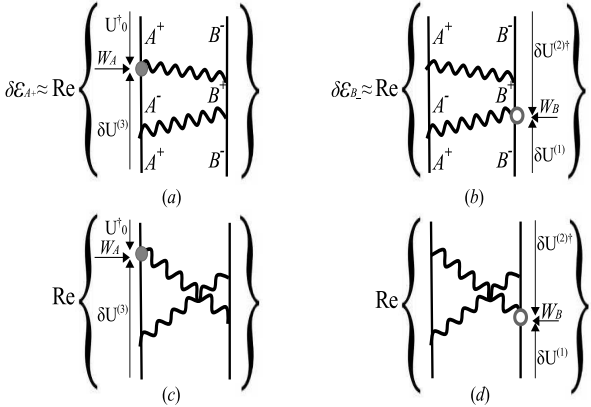}
\caption{Diagrammatic representation of the single-atom level shifts $\delta\mathcal{E}_{A_{+}}=\hbar\delta\omega_{A}$ --Fig.$(a)$-- and $\delta\mathcal{E}_{B_{-}}=-\hbar\delta\omega_{B}/2$ --Fig.$(b)$-- in the
quasiresonant approximation. Diagrams $(c)$ and $(d)$ depict the resonant contributions to  $\delta\mathcal{E}_{A_{+}}$ and  $\delta\mathcal{E}_{B_{-}}$ respectively, beyond the quasiresonant approximation.}\label{figure3}
\end{figure}

 As for the shift $\delta\langle\dot{\sigma}_{A}(T)\rangle$, the equations (7,49) of Ref.\cite{Milonni} read, in terms of Schr\"odinger's operators,
\begin{align}
\delta\langle\dot{\sigma}_{A}&(T)\rangle^{qr}\simeq i\hbar^{-1}\langle\sigma_{A}(T-2R/c)\rangle\nonumber\\
&\times\langle\Psi(0)|\mathbb{U}^{\dagger}_{0}(T)\mathbf{d}_{A}\cdot\mathbf{E}^{(-)}(\mathbf{R}_{A})\delta\mathbb{U}^{(3)}(T)|\Psi(0)\rangle\nonumber\\
&+\langle....\rangle.\label{dA}
\end{align}
In this equation $\mathbb{U}^{\dagger}_{0}(T)\mathbf{E}^{(-)}(\mathbf{R}_{A})\delta\mathbb{U}^{(3)}(T)$ is the Heisenberg field which, at $\mathbf{R}_{A}$ and time $T$, annihilates the photons emitted by dipole $B$, whose dipole moment has been induced by the field emitted by atom $A$ in the first place. It corresponds to the field $\mathcal{E}^{(+)}_{AB}(T)\mu_{A}^{-1}\sigma_{A}^{-1}(T-2R/c)$ in Eq.(48) of Ref.\cite{Milonni}. The relevant component of $\delta\mathbb{U}^{(3)}$ is here proportional to  $W_{A}W_{B}^{2}$ (T-ordered). The corresponding frequency shift, $\delta\omega_{A}$,  is given by the diagram $(a)$ of Fig.\ref{figure3}, which is indeed equivalent to the contributions of Fig.\ref{figure2}($a$) to $\delta \mathcal{E}$ and of Fig.\ref{figure1}($a$) to $\langle W_{A}/2\rangle_{T}$. Therefore, up to the approximations used in Refs.\cite{Berman,Me,Milonni}, the three quantities $\delta\mathcal{E}$, $\langle W_{A}/2\rangle_{T}$ and $\hbar\delta\omega_{A}$ are equivalent. 

As for the shift $\delta\omega_{B}$, Eqs.(13,29) of Ref.\cite{Milonni} for $\delta\langle\dot{\sigma}_{B}(T)\rangle$ read, in terms of Schr\"odinger's operators,
\begin{align}
\delta\langle\dot{\sigma}_{B}&(T)\rangle^{qr}\simeq -2i\hbar^{-1}\langle\sigma_{B}(T)\rangle\nonumber\\
&\times\langle\Psi(0)|\delta\mathbb{U}^{(2)\dagger}(T)\mathbf{d}_{B}\cdot\mathbf{E}^{(-)}(\mathbf{R}_{B})\delta\mathbb{U}^{(1)}(T)|\Psi(0)\rangle\nonumber\\
&+\langle....\rangle.\label{dB}
\end{align}
In this equation $\mathbb{U}^{\dagger}_{0}(T)\mathbf{E}^{(-)}(\mathbf{R}_{B})\delta\mathbb{U}^{(1)}(T)$ is the Heisenberg field which annihilates at atom $B$ and time $T$ the photons emitted by dipole $A$ [cf. Eq.(21) of Ref.\cite{Milonni}], while $\delta\mathbb{U}^{(2)\dagger}(T)\mathbf{d}_{B}\mathbb{U}_{0}(T)$ is the Heisenberg dipole moment of atom $B$, which is induced  by the field emitted by atom $A$. Hence, the relevant component of $\delta\mathbb{U}^{(2)}$ is here proportional to $W_{A}W_{B}$ (T-ordered),  while that of  $\delta\mathbb{U}^{(1)}$ is proportional to $W_{A}$.  The contribution of Eq.(\ref{dB}) to $\delta\omega_{B}$ is represented diagrammatically in Fig.\ref{figure3}($b$). Same as for the case of $\delta\omega_{A}$,  due to the symmetry of that diagram  $\hbar\delta\omega_{B}$ equals  $-\langle W_{B}\rangle^{qr}_{T}$. As well remarked by the authors of Ref.\cite{Milonni}, atom $A$  in Eq.(\ref{dB}) remains unaffected by the presence of atom $B$, which explains also the factor $\Theta(T-
R/c)$ in Eq.(\ref{WB}) in the place of  $\Theta(T-2R/c)$ in Eq.(\ref{WA}). Pictorially, this difference can be seen from the diagrams of Fig.\ref{figure3}. In diagram 3$(a)$, reading from bottom up,  a first photon is emitted from $A$ and absorbed by $B$ while a second photon is later emitted from $B$ and absorbed by $A$. Thus, the minimum time required for the two photons to fly between the atoms, one after the other, is $2R/c$. On the contrary, in diagram 3$(b)$ both photons are emitted from atom $A$. This is, reading from top down the photon emitted from $A$ induces a dipole moment in atom $B$; while reading from bottom up, the photon emitted from $A$ is absorbed by dipole $B$. The minimum time required for the two photons to fly from $A$ to $B$ is here $R/c$, since both photons can depart from $A$ at the same time.

Lastly, neglecting the level shifts due to the interaction of the two atoms in their ground states, the authors of Ref.\cite{Milonni} found for the single-atom shifts of the levels $A_{+}$ and $B_{-}$,   $\delta\mathcal{E}_{A_{+}}=\hbar\omega_{A}$ and $\delta\mathcal{E}_{B_{-}}=-\hbar\delta\omega_{B}/2$, i.e.,     $\delta\mathcal{E}_{A_{+}}=\langle W_{A}/2\rangle_{T}$ and $\delta\mathcal{E}_{B_{-}}=\langle W_{B}/2\rangle_{T}$ respectively, whose expressions are those in Eqs.(\ref{WAT}) and (\ref{WBT}). Milonni and Rafsanjani have interpreted the latter quantity as the quadratic Stark shift of level $B_{-}$ under the field induced by the atom $A$ excited, which is certainly the case at our order of approximation, $\mathcal{O}(W^{4})$. Further, they argue that $\delta\mathcal{E}_{B_{-}}$ is the shift which accompanies the irreversible excitation transfer from $A$ to $B$, for which the conditions $\Gamma_{A}T\ll1$, $\Gamma_{B}T\gg1$ are necessary. Although this interpretation is of course possible, I rather 
prefer to interpret $\delta\mathcal{E}_{B_{-}}$ as the quasistationary vdW potential of atom $B$, since that is the quantity  which explicitly appears in Eq.(\ref{dB})  upon averaging  $\langle W_{B}(T)\rangle^{qr}/2$ over time, even for $\Gamma_{A,B}T\ll1$ --see also Sec.\ref{TvdW}.

Beyond the quasiresonant approximation, although not explicitly computed by the authors of Ref.\cite{Milonni}, the resonant contributions to  $\delta\mathcal{E}_{A_{+}}$ and  $\delta\mathcal{E}_{B_{-}}$ are depicted by diagrams \ref{figure3}$(c)$ and  \ref{figure3}$(d)$ respectively. Again, because of the up-down symmetry of both diagrams, their contributions are the same as those of diagrams \ref{figure1}$(c)$ and \ref{figure1}$(o)$ for  $\langle W_{A}\rangle/2$ and  $\langle W_{B}\rangle/2$ respectively. In the adiabatic approximation (i.e., quasistationary), the frequency poles provided by diagrams \ref{figure3}$(a)$ and \ref{figure3}$(c)$ were already given in Eq.(\ref{poleA}). As for the poles provided by the  diagrams  \ref{figure3}$(b)$ and  \ref{figure3}$(d)$ for $\delta\mathcal{E}_{B_{-}}$--equivalently, by diagrams \ref{figure1}$(m)$ and \ref{figure1}$(o)$ for $\langle W_{B}/2\rangle_{T}$--  Eqs.(\ref{diagm}) and(\ref{diago}) yield
\begin{eqnarray}
&c&/[\/\Delta_{AB}(k-k_{A}-i\eta/c)(k'-k_{A}+i\eta/c)]\quad\textrm{and}\label{polB}\\
&-&c/[(k+k'-\/\Delta_{AB})(k-k_{A}-i\eta/c)(k'-k_{A}+i\eta/c)],\nonumber
\end{eqnarray}
respectively, with $\eta\rightarrow0^{+}$. In contrast to the complex poles of Eq.(\ref{poleA}), the signs of the imaginary parts of the two poles in Eq.(\ref{polB}) differ. This results in the opposite contribution of the term proportional to Im$^{2}\mathbb{G}^{(0)}(\mathbf{R},\omega_{A})$ in Eq.(\ref{WBT}) with respect to that in Eq.(\ref{WAT}); hence, the aforementioned monotonic form of $\langle W_{B}/2\rangle_{T}$.

From this analysis I conclude that the results of Ref.\cite{Milonni}  are fully equivalent, in the quasistationary approximation, to those of Refs.\cite{Berman} and \cite{Me} together with the results of Sec.\ref{Mer}, even though they refer to different physical quantities.

\section{ vdW potentials vs. rate of phase-shift in the interaction between two identical atoms excited}\label{Rydberg}

\begin{figure}[h]
\includegraphics[height=6.5cm,width=7.9cm,clip]{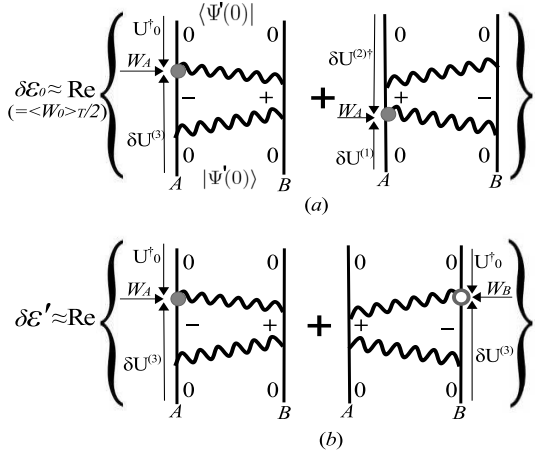}
\caption{Diagrammatic representation of the single-atom level shift $\delta\mathcal{E}_{0}$ --Fig.$(a)$-- and the two-atom level shift $\delta\mathcal{E}'$ --Fig.$(b)$-- for the interaction of two identical excited atoms in the quasiresonant approximation.}\label{figure4}
\end{figure}

I show in this section that the aforemention equivalence between the phase-shift rate of the two-atom state, the vdW potential and the level shift of the excited atom does not hold generally. Let us take as a counterexample the interaction between two identical three-level atoms. Let us assume that the states of each atom are  connected through consecutive E1 transitions, $|-\rangle\rightarrow_{E_{1}}|0\rangle\rightarrow_{E_{1}}|+\rangle$, and let us consider as initial state that in which both atoms, $A$ and $B$, are  placed in their intermediate states, $|\Psi'(0)\rangle=|0\rangle_{A}\otimes|0\rangle_{B}\otimes|0_{\gamma}\rangle$.  The expressions for the phase-shift rate and for the  vdW potential of each atom are in this case $\delta\mathcal{E}'=\textrm{Re}\{\langle\Psi'(0)|\mathbb{U}_{0}^{\dagger}(T)(W_{A}+W_{B})\mathbb{U}(T)|\Psi'(0)\rangle_{T}\}$ and  $\langle W_{0}/2\rangle_{T}\equiv\langle W_{A,B}/2\rangle_{T}=\langle\Psi'(0)|\mathbb{U}^{\dagger}(T)W_{A,B}\mathbb{U}(T)|\Psi'(0)\rangle_{T}/2$, 
respectively.
 In addition, let us
consider that quasiresonant conditions meet, meaning here that  $\Delta_{+-}\equiv(\omega_{+}-\omega_{0})-(\omega_{0}-\omega_{-})$ is such that $|\Delta_{+-}|\ll(\omega_{+}-\omega_{0}),(\omega_{0}-\omega_{-})$. This situation is of particular interest, as it corresponds to the binary interaction between identical circular Rydberg atoms \cite{Haroche}.
It is easy to verify in this example that, while the one-atom shift of the level $0$, $\delta\mathcal{E}_{0}$, still equals the vdW potential, the phase-shift rate of the two-atom state is not equivalent. In Figs.\ref{figure4}($a$) and  ($b$) I depict diagrammatically the processes which contribute to $\delta\mathcal{E}_{0}=\langle W_{0}/2\rangle_{T}$ and $\delta\mathcal{E}'$ in the quasistationary approximation. They yield --see Eqs.(\ref{ER0qr}) and (\ref{Epqr}) in the Appendix--
\begin{align}
\delta\mathcal{E}^{qr}_{0}&=\langle W_{0}/2\rangle^{qr}_{T}=\frac{-2(\omega_{0}-\omega_{-})^{4}}{\epsilon_{0}^{2}\hbar c^{4}\Delta_{+-}}\mu^{i}_{-}\mu^{j}_{+}\mu^{p}_{-}\mu^{q}_{+}\label{WARydb}\\
&\times\textrm{Re}\:G_{ij}^{(0)}[\mathbf{R},(\omega_{0}-\omega_{-})]\textrm{Re}\:G_{pq}^{(0)}[\mathbf{R},(\omega_{0}-\omega_{-})],\nonumber\\
\delta\mathcal{E}^{'qr}&=\frac{-2(\omega_{0}-\omega_{-})^{4}}{\epsilon_{0}^{2}\hbar c^{4}\Delta_{+-}}\mu^{i}_{-}\mu^{j}_{+}\mu^{p}_{-}\mu^{q}_{+}\label{dERydb}\\
&\times\Bigl\{\textrm{Re}\:G_{ij}^{(0)}[\mathbf{R},(\omega_{0}-\omega_{-})]\textrm{Re}\:G_{pq}^{(0)}[\mathbf{R},(\omega_{0}-\omega_{-})]\nonumber\\
&-\textrm{Im}\:G_{ij}^{(0)}[\mathbf{R},(\omega_{0}-\omega_{-})]\textrm{Im}\:G_{pq}^{(0)}[\mathbf{R},(\omega_{0}-\omega_{-})]\Bigr\},\nonumber
\end{align}
where  $\mu^{i}_{-}=\langle0|d_{A,B}^{i}|-\rangle$, $\mu^{j}_{+}=\langle0|d_{A,B}^{j}|+\rangle$.
In the far field Eqs.(\ref{WARydb}) and (\ref{dERydb}) approach, respectively,
\begin{eqnarray}
\delta\mathcal{E}^{qr}_{0}&\simeq&\frac{\mathcal{U}'_{ijpq}}{R^{2}}\mu^{i}_{-}\mu^{j}_{+}\mu^{p}_{-}\mu^{q}_{+}\cos^{2}{[R(\omega_{0}-\omega_{-})/c]},\nonumber\\
\delta\mathcal{E}^{'qr}&\simeq&\frac{\mathcal{U}'_{ijpq}}{R^{2}}\mu^{i}_{-}\mu^{j}_{+}\mu^{p}_{-}\mu^{q}_{+}\cos{[2R(\omega_{0}-\omega_{-})/c]},\nonumber
\end{eqnarray}
where $\mathcal{U}'_{ijpq}=\frac{-2(\omega_{0}-\omega_{-})^{4}}{(4\pi\epsilon_{0})^{2}\hbar c^{4}\Delta_{+-}}\alpha_{ij}\alpha_{pq}$. This time both quantities oscillate in space, but in a different manner.

\section{Dynamical vdW potentials}\label{TvdW}

In Sec.\ref{3papers} I have discussed the equivalence between the energies computed in Refs.\cite{Berman}, \cite{Me} and \cite{Milonni} for the case of the two-atom interaction with  one atom in an excited state. In principle, all those approaches consider that atom $A$ is initially excited, with the atoms in the unentangled state, $|\Psi(0)\rangle=|A_{+}\rangle\otimes|B_{-}\rangle\otimes|0_{\gamma}\rangle$. As explained earlier, they obtain  equivalent stationary values out of their time-dependent calculations, either by discarding rapidly oscillating terms \cite{Berman}, by invoking the adiabatic   switching of the interaction $W$ \cite{Me}, or by averaging over time scales greater than $|\Delta_{AB}^{-1}|$ \cite{Me,Milonni}. Only Berman, in the Appendix of Ref.\cite{Berman},  adopts a more realistic setup by considering that the excitation of atom $A$ is adiabatic with respect to the time scale $|\Delta_{AB}^{-1}|$. That is, he considers that the duration of a $\pi$ pulse resonant with the 
transition of atom $A$, $\tau$, is such that $|\Delta_{AB}^{-1}|\ll\tau\ll T$. In turn, this procedure has the same effect as the adiabatic approximation of Ref.\cite{Me} or the time average of Ref.\cite{Milonni}, as it allows him to get rid off the rapidly oscillating terms in the two-atom wave function, $\tilde{b}_{A}(T)=\langle\tilde{\Psi}(0)|\tilde{\Psi}_{I}(T)\rangle$, where this time $|\tilde{\Psi}(0)\rangle=|A_{-}\rangle\otimes|B_{-}\rangle\otimes|0_{\gamma}\rangle$ and the interaction of the atoms in their ground states is considered negligible.

In the opposite limit, the fully time-dependent result of Ref.\cite{Me} for $\langle W_{A}(T)\rangle/2$ relies on the sudden excitation of atom $A$. In this case, the validity of Eq.(4) in Ref.\cite{Me} not only requires that the temporal resolution be less than $|\Delta_{AB}^{-1}|$, but also that $\tau\ll|\Delta_{AB}^{-1}|$ in order to neglect the effect of $W$ within  the pulse.

\begin{figure}[h]
\includegraphics[height=4.4cm,width=8.9cm,clip]{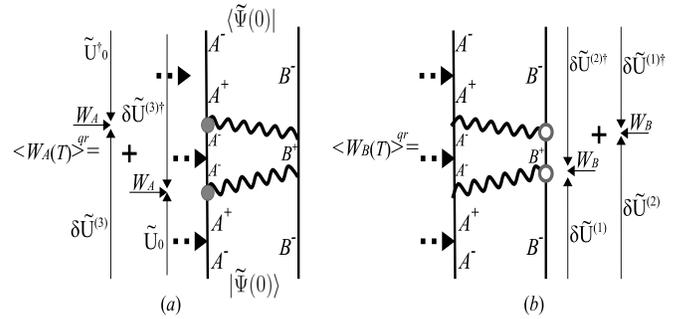}
\caption{Diagrammatic representation of the vdW potentials under the action of a $\pi$ pulse in the quasiresonant approximation. Thick dashed arrows represent the action of the pulse, and tilded propagators incorporate its action in the states of atom $A$.}\label{figure5}
\end{figure}

In the following I consider the case in which the excitation of atom $A$ is neither sudden nor adiabatic, but driven by a $\pi$ pulse of frequency $\Omega$, resonant with the transition $|A_{-}\rangle\rightarrow|A_{+}\rangle$, and such that $\Omega\ll\omega_{A}$. Note that in this case only the vdW potentials make physical sense since the interaction becomes dynamical. In its simplest form, the interaction of the pulse with the atom $A$ is given by the  Hamiltonian
\begin{equation}
H_{R}(t)=\frac{\hbar\Omega}{2}e^{i\omega_{A}t}|A_{-}\rangle\langle A_{+}|+\textrm{h.c.},\quad0\leq t\leq\pi/\Omega.
\end{equation}
In addition, I incorporate the effect of free-space spontaneous emission in a Weisskopf-Wigner approximation, and consider negligible the vdW interaction between the atoms in their ground state.
The calculation of $\langle W_{A}(T)\rangle/2$ is analogous to that carried out in Ref.\cite{Me}, with the differences that the initial state is now $|\tilde{\Psi}(0)\rangle=|A_{-}\rangle\otimes|B_{-}\rangle\otimes|0_{\gamma}\rangle$ and the  unperturbed time-propagator components of the states of atom $A$ read \cite{PRACoh}
\begin{align}
&\cos{(\Omega t/2)}|A_{-}\rangle\langle A_{-}|,\quad e^{-i(\omega_{A}-i\Gamma_{A}/2)t}\cos{(\Omega t/2)}|A_{+}\rangle\langle A_{+}|,\nonumber\\
-i&\sin{(\Omega t/2)}|A_{-}\rangle\langle A_{+}|,\quad -ie^{-i(\omega_{A}-i\Gamma_{A}/2)t}\sin{(\Omega t/2)}|A_{+}\rangle\langle A_{-}|,\nonumber
\end{align}
within the pulse, $0\leq t\leq\pi/\Omega\leq T$; whereas they are  $|A_{-}\rangle\langle A_{-}|$, $e^{-i(\omega_{A}-i\Gamma_{A}/2)t}|A_{+}\rangle\langle A_{+}|$ outside the pulse, $T\geq t>\pi/\Omega$. As for the unperturbed time-propagator components of the states of atom $B$,  they are  $|B_{-}\rangle\langle B_{-}|$, $e^{-i(\omega_{B}-i\Gamma_{B}/2)t}|B_{+}\rangle\langle B_{+}|$, for $0\leq t\leq T$. In this equations I consider $\Gamma_{A,B}\ll\Omega,|\Delta_{AB}|$. Again restricting the calculation to the quasiresonant approximation, at leading order in $\Delta_{AB}/\omega_{A,B}$ the diagram which contributes the most to $\langle W_{A}(T)\rangle^{qr}/2$ is that of Fig.\ref{figure5}($a$), where the action of the pulse is represented by thick dashed arrows and tilded propagators incorporate its action in the states of atom $A$.

An analogous expression to that of Eq.(2) in Ref.\cite{Me} yields, for $T\geq\pi/\Omega$,
\begin{widetext}
\begin{align}
\langle W_{A}(T)\rangle^{qr}/2&=\frac{2\alpha_{f} c^{3}}{\pi\epsilon_{0}e^{2}}\mu_{A}^{i}\mu_{B}^{j}\mu_{B}^{p}\mu_{A}^{q}\int_{-\infty}^{+\infty}\textrm{d}k\:k^{2}\textrm{Im}G^{(0)}_{ij}(kR)\int_{-\infty}^{+\infty}\textrm{d}k'\:k'^{2}\textrm{Im}G^{(0)}_{pq}(k'R)\:\Theta(T-2R/c)\:\Theta(T-\pi/\Omega)\nonumber\\
&\times\Bigl\{\int_{\pi/\Omega}^{T}\textrm{d}t\int_{\pi/\Omega}^{t}\textrm{d}t'\int_{\pi/\Omega}^{t'}\textrm{d}t''+
\Bigl[\int_{\pi/\Omega}^{T}\textrm{d}t\int_{\pi/\Omega}^{t}\textrm{d}t'\int_{0}^{\pi/\Omega}\textrm{d}t''+
\int_{\pi/\Omega}^{T}\textrm{d}t\int^{\pi/\Omega}_{0}\textrm{d}t'\int_{0}^{t'}\textrm{d}t''\nonumber\\
&+\int^{\pi/\Omega}_{0}\textrm{d}t\int^{t}_{0}\textrm{d}t'\int_{0}^{t'}\textrm{d}t''\Bigr]
\sin^{2}{(\Omega t''/2)}\Bigr\}\nonumber\\
&\times\Bigl[\bigl(i\:e^{i(\omega_{A}+i\Gamma_{A}/2)T}e^{-i(T-t)\omega}e^{-i(t-t')(\omega_{B}-i\Gamma_{B}/2)}e^{-i(t'-t'')\omega'}e^{-it''(\omega_{A}-i\Gamma_{A}/2)}\bigr)+(\omega\leftrightarrow\omega')^{*}\Bigr],\nonumber
\end{align}
\end{widetext}
where $\alpha_{f}=e^2/4\pi\epsilon_{0}\hbar c$ is the fine-structure constant and $e$ is the electron charge.
In this equation the first time integral yields the contribution outside the pulse, which is equivalent to the expression of Eq.(4) in Ref.\cite{Me} but for the replacement of $T$ with $T-\pi/\Omega$. The remaining time integrals, which are accompanied by a factor $\sin^{2}{(\Omega t''/2)}$, yield  the contribution within the pulse. Putting altogether one arrives at
\begin{widetext}
\begin{align}\label{beyondad}
\langle W_{A}(T)\rangle^{qr}/2&\simeq(4\pi)^{2}\mathcal{U}^{ijpq}\Bigl\{k_{A}^{4}e^{-\Gamma_{A}T}\Bigl[\textrm{Re}\:G_{ij}^{(0)}(\mathbf{R},\omega_{A})\textrm{Re}\:G_{pq}^{(0)}(\mathbf{R},\omega_{A})-\textrm{Im}\:G_{ij}^{(0)}(\mathbf{R},\omega_{A})\textrm{Im}\:G_{pq}^{(0)}(\mathbf{R},\omega_{A})\Bigr]\nonumber\\
&+\frac{k_{B}^{4}\Omega^{2}e^{-(\Gamma_{A}+\Gamma_{B})T/2}}{2(\Delta_{AB}^{2}-\Omega^{2})}\Bigl[\Bigl(\textrm{Re}\:G_{ij}^{(0)}(\mathbf{R},\omega_{B})]\textrm{Re}\:G_{pq}^{(0)}(\mathbf{R},\omega_{B})-\textrm{Im}\:G_{ij}^{(0)}(\mathbf{R},\omega_{B})\textrm{Im}\:G_{pq}^{(0)}(\mathbf{R},\omega_{B})\Bigr)\nonumber\\
&\times\bigl(\cos{(\Delta_{AB}T)}+\cos{[\Delta_{AB}(T-\pi/\Omega)]}\bigr)-2\textrm{Re}\:G_{ij}^{(0)}(\mathbf{R},\omega_{B})\textrm{Im}\:G_{pq}^{(0)}(\mathbf{R},\omega_{B})\nonumber\\
&\times\bigl(\sin{(\Delta_{AB}T)}+\sin{[\Delta_{AB}(T-\pi/\Omega)]}\bigr)\Bigl]\Bigr\}.
\end{align}
\end{widetext}
From this expression we see that, as anticipated by Berman \cite{Berman}, only the terms proportional to $k_{A}^{4}$, which equal Eq.(\ref{WAT}), survive an adiabatic excitation with $\Omega/\Delta_{AB}\rightarrow0$. Leading order corrections are  $\mathcal{O}[(\Omega/\Delta_{AB})^{2}]$. It is also only those terms that survive the limits $\Gamma_{B}T\gg1$, $\Gamma_{A}T\ll1$, and regardless of the ratio $\Omega/\Delta_{AB}$. In the opposite limit, i.e., for a sudden excitation with $\Delta_{AB}/\Omega\rightarrow0$, Eq.(\ref{beyondad}) equals Eq.(4) of Ref.\cite{Me} in the limit $\Gamma_{A,B}T\ll1$. Another interesting situation is that of a resonant excitation with $|\Omega/\Delta_{AB}|\rightarrow1$,  in which case one obtains
\begin{widetext}
\begin{align}\label{beyondadres}
\langle W_{A}(T)\rangle^{qr}/2&\simeq(4\pi)^{2}\mathcal{U}^{ijpq}\Bigl\{k_{A}^{4}e^{-\Gamma_{A}T}\Bigl[\textrm{Re}\:G_{ij}^{(0)}(\mathbf{R},\omega_{A})\textrm{Re}\:G_{pq}^{(0)}(\mathbf{R},\omega_{A})-\textrm{Im}\:G_{ij}^{(0)}(\mathbf{R},\omega_{A})\textrm{Im}\:G_{pq}^{(0)}(\mathbf{R},\omega_{A})\Bigr]\nonumber\\
&-\frac{\pi k_{B}^{4}e^{-(\Gamma_{A}+\Gamma_{B})T/2}}{4}\Bigl[\Bigl(\textrm{Re}\:G_{ij}^{(0)}(\mathbf{R},\omega_{B})]\textrm{Re}\:G_{pq}^{(0)}(\mathbf{R},\omega_{B})-\textrm{Im}\:G_{ij}^{(0)}(\mathbf{R},\omega_{B})\textrm{Im}\:G_{pq}^{(0)}(\mathbf{R},\omega_{B})\Bigr)\sin{(\Omega T)}\nonumber\\
&+2\textrm{Re}\:G_{ij}^{(0)}(\mathbf{R},\omega_{B})\textrm{Im}\:G_{pq}^{(0)}(\mathbf{R},\omega_{B})\cos{(\Omega T)}\Bigl]\Bigr\},\qquad|\Omega/\Delta_{AB}|\rightarrow1.
\end{align}

As for the vdW potential of atom $B$, a similar calculation yields for $T\geq\pi/\Omega$ --see Eq.(\ref{beyondadBb}) and Fig.\ref{figure5}($b$),
\begin{align}\label{beyondadB}
\langle W_{B}(T)\rangle^{qr}/2&\simeq(4\pi)^{2}\mathcal{U}^{ijpq}\Bigl\{k_{A}^{4}e^{-\Gamma_{A}T}\Bigl[\textrm{Re}\:G_{ij}^{(0)}(\mathbf{R},\omega_{A})\textrm{Re}\:G_{pq}^{(0)}(\mathbf{R},\omega_{A})+\textrm{Im}\:G_{ij}^{(0)}(\mathbf{R},\omega_{A})\textrm{Im}\:G_{pq}^{(0)}(\mathbf{R},\omega_{A})\Bigr]\nonumber\\
&+\frac{k_{A}^{2}k_{B}^{2}\Omega^{2}e^{-(\Gamma_{A}+\Gamma_{B})T/2}}{2(\Delta_{AB}^{2}-\Omega^{2})}\Bigl[\Bigl(\textrm{Re}\:G_{ij}^{(0)}(\mathbf{R},\omega_{A})\textrm{Re}\:G_{pq}^{(0)}(\mathbf{R},\omega_{B})+\textrm{Im}\:G_{ij}^{(0)}(\mathbf{R},\omega_{A})\textrm{Im}\:G_{pq}^{(0)}(\mathbf{R},\omega_{B})\Bigr)\nonumber\\
&\times\bigl(\cos{(\Delta_{AB}T)}+\cos{[\Delta_{AB}(T-\pi/\Omega)]}\bigr)
-\Bigl(\textrm{Re}\:G_{ij}^{(0)}(\mathbf{R},\omega_{A})\textrm{Im}\:G_{pq}^{(0)}(\mathbf{R},\omega_{B})\nonumber\\
&-\textrm{Im}\:G_{ij}^{(0)}(\mathbf{R},\omega_{A})\textrm{Re}\:G_{pq}^{(0)}(\mathbf{R},\omega_{B})\Bigr)\bigl(\sin{(\Delta_{AB}T)}+\sin{[\Delta_{AB}(T-\pi/\Omega)]}\bigr)\Bigl]\Bigr\}.
\end{align}
\end{widetext}
Again, only the  terms proportional to $k_{A}^{4}$, which equal Eq.(\ref{WBT}), survive the limit $\Omega/\Delta_{AB}\rightarrow0$, regardless of the value of the ratio  $\Gamma_{B}/\Gamma_{A}$. Nonetheless, it is also those terms that survive  the limits $\Gamma_{B}T\gg1$, $\Gamma_{A}T\ll1$, which are necessary for an irreversible excitation transfer, and regardless of the excitation rate.

\section{Discussion}\label{Disc}
I have shown that, for the problem of the interaction between two dissimilar atoms with one atom in an excited state, the expressions for the phase-shift rate of the two-atom state computed by Berman \cite{Berman}, $\delta\mathcal{E}/\hbar$, for the level shift of the excited atom  computed by Milonni \emph{et al.} \cite{Milonni}, $\delta\mathcal{E}_{A_{+}}$, and for the vdW potential of the excited atom  computed by Donaire \emph{et al.}, $\langle W_{A}\rangle_{T}/2$, are  equivalent in the quasistationary approximation.  To this end, I have expressed all these quantities in terms of time propagators and Schr\"odinger operators within the framework of time-dependent perturbation theory. Their diagrammatic representations are given in Figs.\ref{figure1}, \ref{figure2} and \ref{figure3}.  As for the level shift of the ground state atom  computed in Ref.\cite{Milonni}, $\delta\mathcal{E}_{B_{-}}$, I have shown that it is equivalent to its quasistationary vdW potential, 
$\langle W_{B}\rangle_{T}/2$, regardless of the relaxation rate $\Gamma_{B}$ and of the existence of a continuous distribution of final states. Although the latter condition together with $\Gamma_{B}/\Gamma_{A}\gg1$ are  necessary  for an irreversible excitation transfer, Eq.(\ref{WBT}) is equally applicable for $\Gamma_{A,B}T\ll1$, in which case the excitation exchange is reversible. The lack of reciprocity which derives from the inequality $\langle W_{A}\rangle/2\neq\langle W_{B}\rangle/2$ deserves further study which lies outside the scope of the present article.

Beyond the  quasistationary approximation, I have shown the dependence of the dynamical vdW potentials on the frequency of the excitation pulse, $\Omega$, and on the spontaneous emission rates, $\Gamma_{A,B}$. As anticipated by Berman \cite{Berman}, the quasistationary results are recovered for $\Gamma_{A}T\ll1$ in the adiabatic limit, $\Delta_{AB}/\Omega\rightarrow0$. Same results are obtained for  $\Gamma_{B}T\gg1$, regardless of the ratio $\Delta_{AB}/\Omega$. However, the dynamical vdW potentials depend generally on the manner the atom $A$ was excited, as given by Eqs.(\ref{beyondad}) and (\ref{beyondadB}).

I conclude that, in agreement with Berman \cite{Berman}, Milonni and Rafsanjani \cite{Milonni}, the reason for the discrepancy between the different expressions of the two-atom interaction energy in precedent works is that they refer to different physical quantities. In particular, while the van der Waals potential of the excited atom oscillates in space, the van der Waals potential of the ground state atom presents a monotonic form.

The equivalence between the phase-shift rate of the two-atom wave function and the quasistationary vdW potential of an excited atom is however not generic. This point is illustrated  in Sec.\ref{Rydberg} where, for the problem of the interaction between two identical atoms excited,  it is shown that $\delta\mathcal{E}'\neq\langle W_{0}/2\rangle_{T}$.

Concerning the experimental observation of the quantities computed in Refs.\cite{Berman,Me,Milonni}, the vdW potentials $\langle W_{A,B}(T)\rangle/2$ can be observed  through the vdW forces experienced by each atom when placed inside harmonic traps, which are  proportional to the displacements of the atoms with respect to their equilibrium positions in the absence of interaction. On the other hand, the phase-shift of the two-atom state can be observed using atom interferometry. In particular, it is the shift $\delta\mathcal{E}'$ calculated in Sec.\ref{Rydberg} that is observed in the binary interaction of Rydberg atoms through the measurement of population probabilities \cite{Haroche,Rydberg}.
In either case, forces and phase-shifts make reference to the dynamics of the atomic degrees of freedom. In addition to these observables, the frequency of the  photon emitted at $T\rightarrow\infty$ may serve also to quantify the two-atom interaction by spectroscopic means. If no other dissipative channels exist and the kinetic energy associated to the atomic recoil is negligible,  the energy of the emitted photon must equal the energy supplied by the excitation pulse of Sec.\ref{TvdW}. The computation of this quantity is left for a separate publication.

\acknowledgments
I thank M.-P. Gorza for useful discussions on this problem. Financial support from the contracts ANR-10-IDEX-0001-02-PSL and ANR-13-BS04--0003-02, within the group of Quantum Fluctuations and Relativity of the \emph{Laboratoire Kastler-Brossel}, is gratefully acknowledged.

\appendix
\begin{widetext}
\section{Integral expressions for Eqs.(\ref{WB}),  (\ref{poleA}),  (\ref{polB}), (\ref{WARydb}),  (\ref{dERydb})  and (\ref{beyondadB})}\label{Appa}
In this Appendix I write the integral expressions for $\langle W_{B}(T)\rangle^{qr}/2$, $\delta\mathcal{E}^{qr}_{0}$ and $\delta\mathcal{E}^{'qr}$, as well as the time integrals from which the poles in Eqs.(\ref{poleA}) and (\ref{polB}) derive.

As for $\langle W_{B}(T)/2\rangle^{qr}$ in Eq.(\ref{WB}), it reads for a sudden excitation and $\Gamma_{A,B}T\ll1$,
\begin{eqnarray}
\langle W_{B}(T)\rangle^{qr}/2&=&\textrm{Re}\frac{2\alpha_{f} c^{3}}{\pi\epsilon_{0}e^{2}}\mu_{A}^{i}\mu_{B}^{j}\mu_{B}^{p}\mu_{A}^{q}\int_{-\infty}^{+\infty}\textrm{d}k\:k^{2}\textrm{Im}G^{(0)}_{ij}(kR)\int_{-\infty}^{+\infty}\textrm{d}k'\:k'^{2}\textrm{Im}G^{(0)}_{pq}(k'R)\:\Theta(T-R/c)\nonumber\\
&\times&\int_{0}^{T}\textrm{d}t\int_{0}^{t}\textrm{d}t'\int_{0}^{T}\textrm{d}t''\left[\left(-ie^{i(T-t)\omega_{B}}e^{i(t-t')\omega}e^{i\omega_{A}t'}\:e^{-i(T-t'')\omega'}e^{-it''\omega_{A}}\right)+(\omega\leftrightarrow\omega')\right].\label{ECUACIONFORWBT}
\end{eqnarray}
In the case of an excitation driven by a $\pi$ pulse of frequency $\Omega$ the expression for $\langle W_{B}(T)/2\rangle^{qr}$ is that in  Eq.(\ref{beyondadB}), which is computed from the integral
\begin{eqnarray}
\langle W_{B}(T)\rangle^{qr}/2&=&\textrm{Re}\frac{2\alpha_{f} c^{3}}{\pi\epsilon_{0}e^{2}}\mu_{A}^{i}\mu_{B}^{j}\mu_{B}^{p}\mu_{A}^{q}\int_{-\infty}^{+\infty}\textrm{d}k\:k^{2}\textrm{Im}G^{(0)}_{ij}(kR)\int_{-\infty}^{+\infty}\textrm{d}k'\:k'^{2}\textrm{Im}G^{(0)}_{pq}(k'R)\:\Theta(T-R/c)\:\Theta(T-\pi/\Omega)\nonumber\\
&\times&\Bigl\{\Bigl[\int_{\pi/\Omega}^{T}\textrm{d}t\int_{\pi/\Omega}^{t}\textrm{d}t'-i\bigl(\int_{\pi/\Omega}^{T}\textrm{d}t\int^{\pi/\Omega}_{0}\textrm{d}t'+\int^{\pi/\Omega}_{0}\textrm{d}t\int^{t}_{0}\textrm{d}t'\bigr)
\sin^{2}{(\Omega t'/2)}\Bigr]\label{beyondadBb}\\
&\times&\Bigl[\int_{\pi/\Omega}^{T}\textrm{d}t''+i\int^{\pi/\Omega}_{0}\textrm{d}t''
\sin^{2}{(\Omega t''/2)}\Bigr]\Bigr\}\nonumber\\
&\times&\left[\left(-ie^{i(T-t)(\omega_{B}+i\Gamma_{B}/2)}e^{i(t-t')\omega}e^{i(\omega_{A}+i\Gamma_{A}/2)t'}\:e^{-i(T-t'')\omega'}e^{-it''(\omega_{A}-i\Gamma_{A}/2)}\right)+(\omega\leftrightarrow\omega')\right].\nonumber
\end{eqnarray}

The integral expressions which derive from  the diagrams in Figs.\ref{figure4}$(a)$ and \ref{figure4}$(b)$, which yield the results of Eqs.(\ref{WARydb}) and (\ref{dERydb}) respectively, read in the quasistationary (i.e., adiabatic) approximation,
\begin{eqnarray}
\delta\mathcal{E}^{qr}_{0}&=&\langle W_{0}/2\rangle^{qr}_{T}=\textrm{Re}\frac{4\alpha_{f}c^{3}}{\pi\epsilon_{0}e^{2}}\mu_{-}^{i}\mu_{+}^{j}\mu_{+}^{p}\mu_{-}^{q}\int_{-\infty}^{+\infty}\textrm{d}k\:k^{2}\textrm{Im}G^{(0)}_{ij}(kR)\int_{-\infty}^{+\infty}\textrm{d}k'\:k'^{2}\textrm{Im}G^{(0)}_{pq}(k'R)\nonumber\\
&\times&\int_{-\infty}^{T}\textrm{d}t\int_{-\infty}^{t}\textrm{d}t'\int_{-\infty}^{t'}\textrm{d}t''\:e^{\eta(t+t'+t'')}\bigl[i\:e^{2i\omega_{0}T}e^{-i(T-t)(\omega+\omega_{0}+\omega_{-})}e^{-i(t-t')(\omega_{+}+\omega_{-})}e^{-i(t'-t'')(\omega'+\omega_{0}+\omega_{-})}e^{-2it''\omega_{0}}\bigr]\nonumber\\
&+&\textrm{Re}\frac{4\alpha_{f} c^{3}}{\pi\epsilon_{0}e^{2}}\mu_{-}^{i}\mu_{+}^{j}\mu_{+}^{p}\mu_{-}^{q}\int_{-\infty}^{+\infty}\textrm{d}k\:k^{2}\textrm{Im}G^{(0)}_{ij}(kR)\int_{-\infty}^{+\infty}\textrm{d}k'\:k'^{2}\textrm{Im}G^{(0)}_{pq}(k'R)\nonumber\\
&\times&\int_{-\infty}^{T}\textrm{d}t\int_{-\infty}^{t}\textrm{d}t'\int_{-\infty}^{T}\textrm{d}t''\:e^{\eta(t+t'+t'')}\bigl[-ie^{i(T-t)(\omega_{+}+\omega_{-})}e^{i(t-t')(\omega+\omega_{0}+\omega_{-})}e^{2i\omega_{0}t'}\:e^{-i(T-t'')(\omega'+\omega_{0}+\omega_{-})}e^{-2it''\omega_{0}}\bigr],\nonumber\\
&\eta&\rightarrow0^{+},\label{ER0qr}\\
\delta\mathcal{E}^{'qr}&=&2\textrm{Re}\frac{4\alpha_{f}c^{3}}{\pi\epsilon_{0}e^{2}}\mu_{-}^{i}\mu_{+}^{j}\mu_{+}^{p}\mu_{-}^{q}\int_{-\infty}^{+\infty}\textrm{d}k\:k^{2}\textrm{Im}G^{(0)}_{ij}(kR)\int_{-\infty}^{+\infty}\textrm{d}k'\:k'^{2}\textrm{Im}G^{(0)}_{pq}(k'R)\nonumber\\
&\times&\int_{-\infty}^{T}\textrm{d}t\int_{-\infty}^{t}\textrm{d}t'\int_{-\infty}^{t'}\textrm{d}t''\:e^{\eta(t+t'+t'')}\bigl[i\:e^{2i\omega_{0}T}e^{-i(T-t)(\omega+\omega_{0}+\omega_{-})}e^{-i(t-t')(\omega_{+}+\omega_{-})}e^{-i(t'-t'')(\omega'+\omega_{0}+\omega_{-})}e^{-2it''\omega_{0}}\bigr],\nonumber\\&\eta&\rightarrow0^{+}.\label{Epqr}
\end{eqnarray}
In Eq.(\ref{ER0qr}) the first term corresponds to the diagram on the right of Fig.\ref{figure4}$(a)$, while the second term corresponds to the diagram on the left. In contrast, the factor 2 in front of the expression on the r.h.s. of Eq.(\ref{Epqr}) stand for the equivalent contribution of the two diagrams in Fig.\ref{figure4}$(b)$.

As for the poles appearing in the second equation of Eq.(\ref{poleA}), which corresponds to the diagrams \ref{figure1}($c$),  \ref{figure2}($c$) or  \ref{figure3}($c$), they derive from the time integral
\begin{equation}\label{diagc}
\int_{-\infty}^{T}\textrm{d}t\int_{-\infty}^{t}\textrm{d}t'\int_{-\infty}^{t'}\textrm{d}t''\:e^{\eta(t+t'+t'')}\left(i\:e^{i\omega_{A}T}e^{-i(T-t)\omega}e^{-i(t-t')(\omega+\omega'+\omega_{B})}e^{-i(t'-t'')\omega'}e^{-it''\omega_{A}}\right),\:\:\eta\rightarrow0^{+}.
\end{equation}
The poles appearing in the first equation of Eq.(\ref{polB}), which corresponds to the diagrams \ref{figure1}($m$) or \ref{figure3}($b$), derive from the time integral
\begin{equation}\label{diagm}
\int_{-\infty}^{T}\textrm{d}t\int_{-\infty}^{t}\textrm{d}t'\int_{-\infty}^{T}\textrm{d}t''\:e^{\eta(t+t'+t'')}\left(-ie^{i(T-t)\omega_{B}}e^{i(t-t')\omega}e^{i\omega_{A}t'}\:e^{-i(T-t'')\omega'}e^{-it''\omega_{A}}\right),\:\:\eta\rightarrow0^{+}.
\end{equation}
Finally, the poles appearing in the second equation of Eq.(\ref{polB}), which corresponds to the diagrams \ref{figure1}($o$) or \ref{figure3}($d$), derive from the time integral
\begin{equation}\label{diago}
\int_{-\infty}^{T}\textrm{d}t\int_{-\infty}^{t}\textrm{d}t'\int_{-\infty}^{T}\textrm{d}t''\:e^{\eta(t+t'+t'')}\left(-ie^{i(T-t)(\omega+\omega'+\omega_{B})}e^{i(t-t')\omega}e^{i\omega_{A}t'}\:e^{-i(T-t'')\omega'}e^{-it''\omega_{A}}\right),\:\:\eta\rightarrow0^{+}.
\end{equation}

\end{widetext}

\end{document}